\documentclass[aip,preprint]{revtex4-1}
\usepackage{graphicx}
\usepackage[section]{placeins}
 \usepackage{float}
  \usepackage{amsmath}

\begin{document}


\title{On the thermodynamics of fermions at any temperature based on parametrized partition function}

\author{Yunuo Xiong}
\email{xiongyunuo@hbpu.edu.cn}
\affiliation{Center for Fundamental Physics and School of Mathematics and Physics, Hubei Polytechnic University, Huangshi 435003, China}

\author{Hongwei Xiong}
\email{xionghongwei@hbpu.edu.cn}

\affiliation{Center for Fundamental Physics and School of Mathematics and Physics, Hubei Polytechnic University, Huangshi 435003, China}


\affiliation{Wilczek Quantum Center, Shanghai Jiao Tong University, Shanghai 200240, China}

\date{\today}

\begin{abstract}

In this work we study the recently developed parametrized partition function formulation and show how we can infer the thermodynamic properties of fermions based on numerical simulation of bosons and distinguishable particles at various temperatures. In particular, we show that in the three dimensional space defined by energy, temperature and the parameter characterizing parametrized partition function, we can map the energies of bosons and distinguishable particles to fermionic energies through constant-energy contours. We apply this idea to both noninteracting and interacting Fermi systems and show it is possible to infer the fermionic energies at all temperatures, thus providing a practical and efficient approach to obtain thermodynamic properties of Fermi systems with numerical simulation. As an example, we present energies and heat capacities for 10 noninteracting fermions and 10 interacting fermions (more fermions are provided in the appendix) and show good agreement with the analytical result for noninteracting case.
\end{abstract}


\maketitle

\section{Introduction}

The ability to simulate fermions is of paramount importance in the field of numerical calculation, early methods to simulate Fermi systems are primarily the Hartree-Fock method, and density functional approach. Those methods have been applied to gain valuable insights into the atomic structure; unfortunately though, they treat the quantum correlation and exchange effects in an approximate manner, and being able to take such effects into account is crucial for realistic many body quantum systems. Later on, a numerically exact method based on the path integral formulation of quantum mechanics was developed, known as path integral Monte Carlo/molecular dynamics \cite{barker,Tuckerman,cazorla}, and it has been successfully applied to extract thermodynamic properties of Bose systems from ab initio simulations \cite{CeperRMP,boninsegni1,boninsegni2,Hirshberg,Deuterium,Xiong,Xiong3,Xiong4,Xiong5}. In principle, path integral Monte Carlo/molecular dynamics takes all quantum effects into account but when we try to apply this methodology to fermions, we encounter an insurmountable difficulty known as fermion sign problem \cite{ceperley,Alex,troyer,loh,lyubartsev,vozn,Science,Wu,Umrigar,Li,Wei,Yao2,HirshbergFermi,DornheimMod,Xiong2}, where the probabilities used for sampling become negative. In this work, we consider the recently developed parametrized path integral formulation \cite{XiongFSP,XiongPara} and propose a scheme to overcome the difficulties associated with the numerical simulation of Fermi systems, hopefully obtaining a method to numerically study the ab initio properties of Fermi systems.

The recently developed parametrized partition function \cite{XiongFSP} provides a scheme to extrapolate the thermodynamics of fermions, from distinguishable particles and bosons; in particular, the energy as a monotonic function of the extrapolation parameter $\xi$. In a previous work \cite{XiongPara}, an attempt was made to infer the thermodynamics of fermions by numerically simulating the parametrized partition function for $\xi\geq 0$ through path integral molecular dynamics (PIMD), and then extrapolate the results to $\xi=-1$ corresponding to fermions; of course, direct simulation for $\xi<0$ is infeasible due to fermion sign problem where the probability distribution in importance sampling becomes negative, rendering any sampling methods inapplicable. This approach worked well for interacting Fermi systems at medium and high temperature, but the extrapolation scheme was shown to be unreliable at low temperature or for noninteracting Fermi systems. In particular, such extrapolation fails completely for the ground state of noninteracting fermions.

In this work, we further study the properties of parametrized partition function and consider the three dimensional phase space defined by energy, temperature, and $\xi$. We show that, instead of trying to infer fermion energies in the two dimensional plane defined by energy and $\xi$ as we did in the extrapolation scheme, the addition of temperature greatly improves credibility for the inference process; that is, within the framework of this new method we demonstrate how we could connect the thermodynamics of fermions with that of bosons and distinguishable particles through constant-energy contours in this three-dimensional space. Moreover, we show theoretically that the contours can be described by parabolic curves for the examples in the text, giving rise to the aforementioned credibility when performing the inference. Our method can be applied to study both the ground state and finite temperature properties of Fermi systems; that is, by obtaining all the information for the $\xi\geq 0$ region, which can be done efficiently, we gained access to the thermodynamic properties for the Fermi system. To test our method, we apply it for both interacting and noninteracting Fermi systems and give results for heat capacity that can be compared against analytical and other results. We also discuss the general method to improve the precision by making fuller use of the extra parameter in the parametrized partition function, in addition to the temperature.

\section{Theory}

The partition function of $N$  particles is
\begin{equation}
Z(\beta)=Tr(e^{-\beta \hat H}).
\end{equation}
Here $\beta=1/k_B T$, with $k_B$ being the Boltzmann constant and $T$ being the system temperature.
The average energy is $E(\beta)=-\partial\ln Z(\beta)/\partial \beta$, with other parameters being fixed.

For $N$ identical particles, we consider the following parametrized partition function \cite{XiongFSP,XiongPara} with a real parameter $\xi$,
\begin{equation}
Z(\xi,\beta)\sim\sum_{p\in S_N}\xi^{N_p}\int d\textbf{r}_1d\textbf{r}_2\cdots d\textbf{r}_N\left<p\{\textbf{r}\}|e^{-\beta \hat H}|\{\textbf{r}\}\right>.
\label{Xipartition}
\end{equation}
Here $\{\textbf{r}\}$ denotes $\{\textbf{r}_1,\cdots,\textbf{r}_N\}$.  $S_N$ represents the set of $N!$ permutation operations denoted by $p$. The factor $\xi^{N_p}$ is due to the exchange effect of identical particles, with $N_p$ a number defined to be the minimum number of times for which pairs of indices must be interchanged in permutation $p$ to recover the original order. In this parametrized partition function,  the quantum statistics parameter $\xi$ interpolates continuously from bosons ($\xi=1$), distinguishable particles ($\xi=0$), to fermions ($\xi=-1$). In addition, we still have
\begin{equation}
E(\xi,\beta)=-\frac{\partial\ln Z(\xi,\beta)}{\partial \beta}.
\label{energyxi}
\end{equation}
In the above calculation, the parameter $\xi$ is fixed to get the energy $E(\xi,\beta)$.

Using $e^{-\beta\hat H}=e^{-\Delta\beta\hat H}\cdots e^{-\Delta\beta\hat H}$ with $\Delta\beta=\beta/P$ and the technique of path integral, the partition function $Z(\xi,\beta)$ with a general parameter $\xi$ can be also mapped as a classical system of interacting ring polymers \cite{XiongFSP,XiongPara}, based on the idea of recursion formula for identical particles \cite{Hirshberg,HirshbergFermi}. The so called exact numerical simulation of the thermodynamics for a quantum system is through this path integral formalism so that $Z(\xi,\beta)$ can be written as the high dimensional integral of all the coordinates of $NP$ beads. Unfortunately, for negative $\xi$, the fermion sign problem makes the direct numerical calculation of the energy for fermions with path integral formalism infeasible under the condition of large particle number or ultra low temperature. The purpose of the present work is to provide an efficient and reliable method to calculate the energy of fermions for any temperature by calculating firstly the energy for $\xi\geq 0$ with PIMD.

\begin{figure}[htbp]
\begin{center}
 \includegraphics[width=0.75\textwidth]{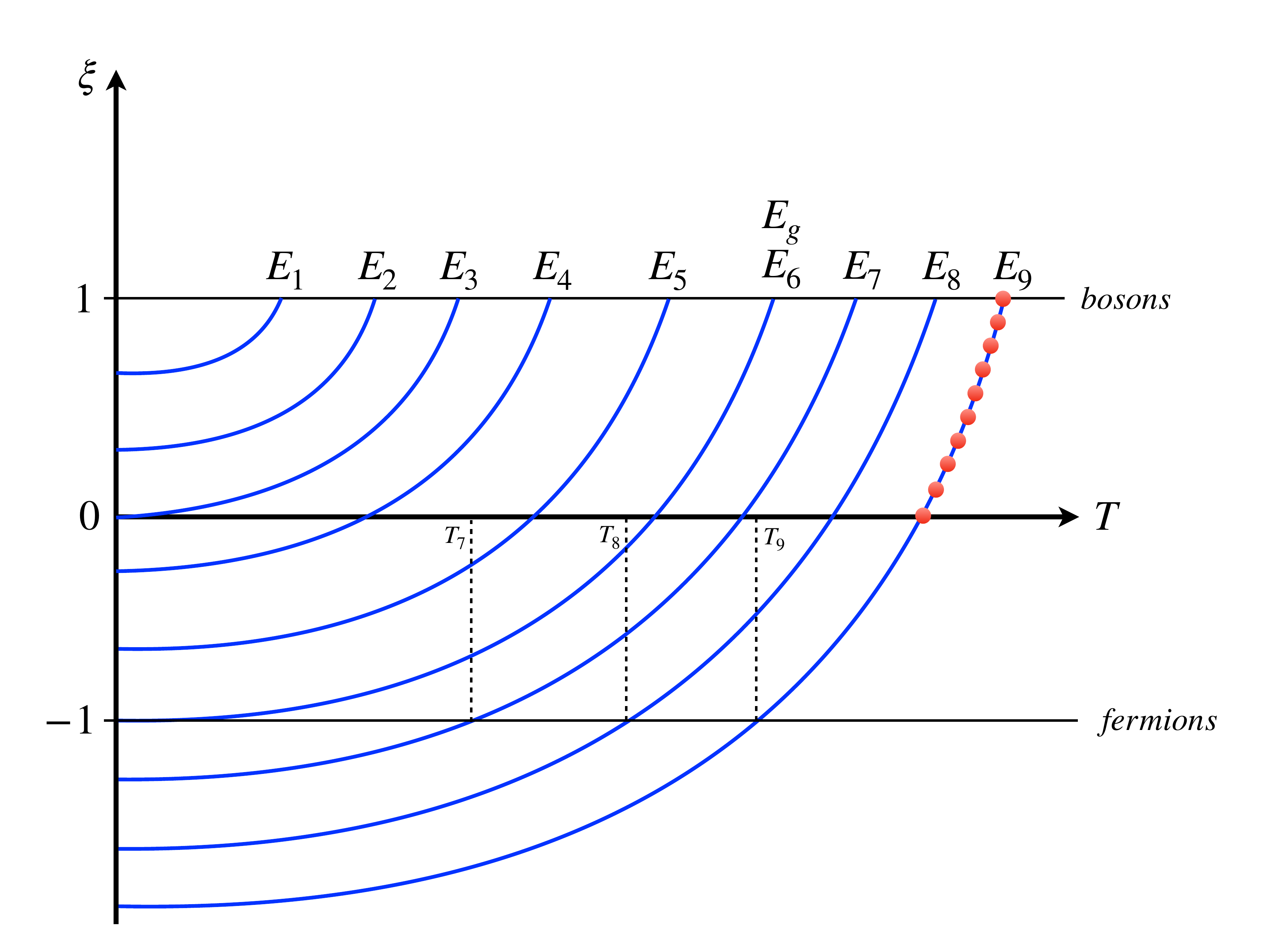} 
\caption{\label{Figillus} A general illustration of the $\xi_E(T)$ curve with constant energy $E$. For the whole region $\xi\geq 0$, we can give exact numerical calculation of the energy $E(\xi,T)$ with path integral molecular dynamics or path integral Monte Carlo, without suffering from fermion sign problem. The general physical consideration tells us that $E(\xi,T)$ has monotonic behavior about the parameters $\xi$ and $T$, so that $E_1<E_2<\cdots <E_9$. The horizontal lines are for bosons with $\xi=1$ and fermions with $\xi=-1$, respectively. As an example, the solid circles give $\xi_E(T)$ for the same energy $E_9$. From the data given by the solid circles, we may determine the function $\xi_E(T)$ based on $\xi_E(T)=a(E)+b(E)T^2$ or more general expression of $\xi_E(T)$. After obtaining $\xi_E(T)$, because $\xi=-1$ for fermions, the vertical dashed line will give the temperature $T_9$ of the fermions with energy $E_9$. The curve for $E_6$ is special because the curve $\xi_E(T)$ passes through the point $(\xi=-1,T=0)$. This means that $E_6$ is the ground state energy of fermions at $T=0$. For $E<E_6$, there is no real number solution of the temperature for $\xi=-1$. This result is not surprising because the energy is already below the ground state energy of fermions.}
\end{center}
\end{figure}

Because $\xi$ is a quantum statistics parameter which has an equivalent repulsive exchange interaction for fermions and equivalent attractive exchange interaction for bosons, for the same temperature, the energy should decrease with increasing $\xi$. In addition, for the same $\xi$, the energy increases with increasing $T$ if other parameters are fixed for most physical systems because of the positive heat capacity. In Fig. \ref{Figillus}, we illustrate a series of contour lines $\xi_E(T)$ with constant energy $E$ which satisfy these two monotonic behavior. In the caption of this figure, we give the method to calculate the temperature of fermions corresponding to a given energy, if we know in advance the property of the contour line $\xi_E(T)$ for any given energy larger than the ground state energy of fermions.

For a given energy, we first consider the behavior of $\xi_E(T)$ near $T=0$. Because the energy is a function of $\xi$ and $\beta$, we also have $\xi(E,\beta)$ and $\beta(E,\xi)$. In this case, we always have the following exact relation based on calculus:
\begin{equation}
\frac{\partial \xi(E,T)}{\partial T}=-\frac{\partial E(\xi,T)/\partial T}{\partial E(\xi,T)/\partial \xi}.
\label{ppp}
\end{equation}

Because $E(\xi,T=0)$ is the ground state energy, while $\partial E(\xi,T)/\partial T$ is the heat capacity, we have $\lim_{T\rightarrow 0}\partial E(\xi,T)/\partial T=0$.
Hence, we obtain the following simple relation:
\begin{equation}
\left.\frac{\partial\xi(E,T)}{\partial T}\right|_{T=0}=0.
\end{equation}
This means that for a given energy, if we expand $\xi_E(T)$ about $T$ with Taylor series, there should be an absence of the linear term. In this case, we have
\begin{equation}
\xi_E(T)=a(E)+b(E)T^2+\sum_{n>2}c_n(E)T^n.
\label{xirelation}
\end{equation}

To consider the behavior of $c_n(E)$ for $n>2$, we discuss $\xi(E,T)$ ($\equiv\xi_E(T)$) at high temperature. For $\beta\rightarrow 0$ (or $T\rightarrow \infty$), $E(\beta,\xi)$ is independent of the quantum statistics parameter $\xi$. In this case, we have
\begin{equation}
E(\xi,\beta\rightarrow 0)=const.
\end{equation}
For small $\beta$, based on the above result and the monotonic behavior of $E(\xi,\beta)$ about $\xi$, for fixed $\beta$, $E(\xi,\beta)$ can be approximated as a linear function of $\xi$, i.e.
\begin{equation}
E(\xi,\beta)\approx E(\xi=0,\beta)+d(\beta)\xi.
\end{equation}
This linear behavior at high temperature was firstly found by the numerical simulation in Ref. \cite{XiongPara}. 

The combination of the low-temperature and high-temperature behaviors suggest that $\xi_E(T)$ may take the following simple expression (see more details in the appendix):
\begin{equation}
\xi_E(T)\approx a(E)+b(E)T^2.
\label{RelationXiT}
\end{equation}
Of course, all the difficulties are now contained in the coefficients $a(E)$ and $b(E)$. Fortunately,  because the above relation is obtained from the general consideration of arbitrary value of $\xi$, it applies to $\xi\geq 0$ too. Hence, the accurate calculation of the energy for $\xi\geq 0$ provides the chance to determine $a(E)$ and $b(E)$, which then predicts the temperature of fermions having energy $E$, by solving $-1=a(E)+b(E) T^2$.

\section{Results}

Now we turn to consider the validity of the above relation with $10$ noninteracting particles in a two-dimensional harmonic trap with potential function $\frac{1}{2}m\omega^2(x^2+y^2)$. The choice of noninteracting particles is due to the fact it provides a standard to test any new method \cite{HirshbergFermi}. In all our calculations, we will use the convention of $\hbar=m=k_B=\omega=1$. For noninteracting particles, we have shown in Ref. \cite{XiongPara} that the energy can be calculated accurately with the following equation in the grand canonical ensemble.
\begin{equation}
N=\sum_{\textbf n}\frac{1}{e^{\beta({\epsilon({\textbf n})-\mu})}-\xi},
\label{chemical}
\end{equation}

\begin{equation}
E(\xi,\beta,N)=\sum_{\textbf n}\frac{\epsilon(\textbf n)}{e^{\beta({\epsilon({\textbf n})-\mu})}-\xi}.
\label{energyc}
\end{equation}
$\epsilon({\textbf n})$ is the single-particle eigenenergies of the system. With given parameters of $N,\beta$, and $\xi$,  from Eq. (\ref{chemical}), we can get the chemical potential $\mu(\xi,\beta,N)$. Using further Eq. (\ref{energyc}), we can get the energy $E(\xi,\beta,N)$ in the grand canonical ensemble.

\begin{figure}[htbp]
\begin{center}
 \includegraphics[width=0.75\textwidth]{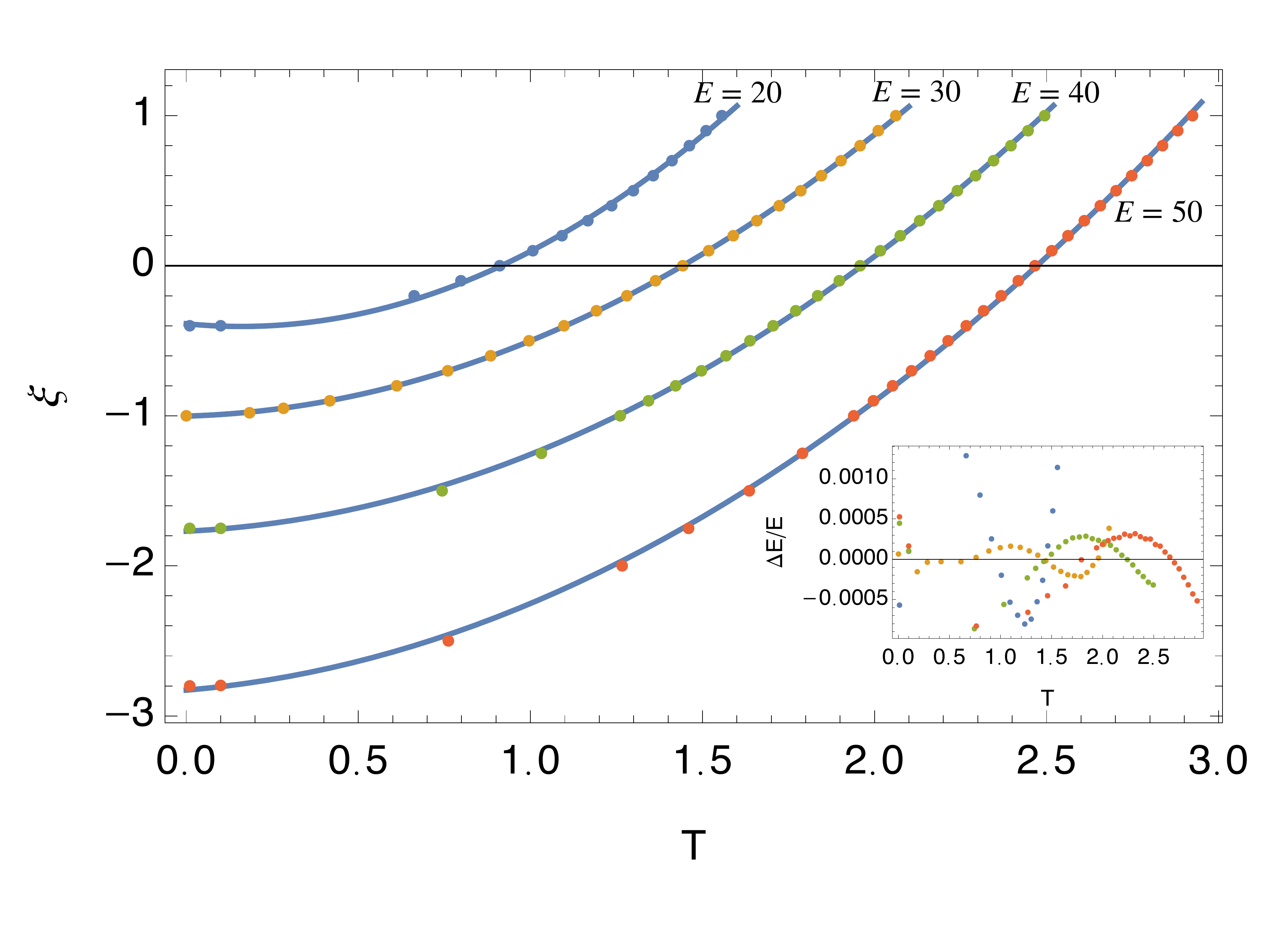} 
\caption{\label{10idealxi} For 10 noninteracting particles, the circles show $\xi$ for different temperatures, while the solid lines are the fitting with the function $\xi_E(T)=a(E)+b(E)T^2$, respectively. Here, four curves are for the contour lines with different energies. In the inset, we show the relative deviation, compared with the parabolic function.}
\end{center}
\end{figure}

In Fig. \ref{10idealxi}, for different energies, the circles give the relation between $\xi$ and $T$. The solid lines are the fitting with Eq. (\ref{RelationXiT}), which shows good agreement. 
In the inset of Fig. \ref{10idealxi}, we show the relative deviation, compared with the parabolic function. The maximum relative deviation is about $0.1\%$, while the mean relative deviation is smaller than $0.03\%$, which does show agreement with the parabolic function.
The validity of the 
relation (\ref{RelationXiT}) paves the way to give us a simple way to predict the energy of fermions for different temperature including the situation of extreme zero temperature. For a given energy $E$, if we know the corresponding temperature $T_0$ for $\xi=0$ and $T_1$ for $\xi=1$, we may get the coefficients $a(E)$ and $b(E)$ based on Eq. (\ref{RelationXiT}). By setting $\xi=-1$ in Eq. (\ref{RelationXiT}), we then get the temperature $T$ for the fermions with this energy. Changing the energy and repeating these simple calculations, we may get continuously the relation between energy and temperature of fermions from zero temperature to high temperature.

\begin{figure}[htbp]
\begin{center}
 \includegraphics[width=0.75\textwidth]{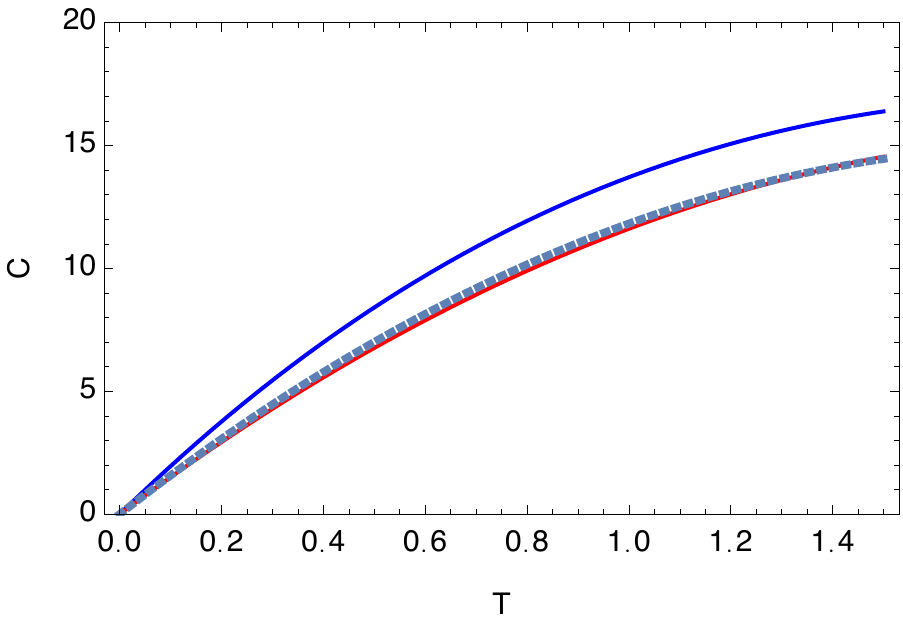} 
\caption{\label{heat} The red solid line is the simulated heat capacity $C(T)$ with our method for $10$ noninteracting fermions, while the blue dashed line is the result based on the formula of grand canonical ensemble. The blue solid line is the simulated heat capacity with our method for $10$ fermions with Coulomb repulsive interaction. In the limit of $T\rightarrow \infty$, the heat capacity should be $20$ for noninteracting case.}
\end{center}
\end{figure}

We still consider the example of $10$ noninteracting particles. For both $\xi=0$ and $\xi=1$, we get a series of energies for different temperatures, based on Eqs. (\ref{chemical}) and (\ref{energyc}). Of course, we may also calculate these data accurately by PIMD \cite{XiongPara}. Here, we use Eqs. (\ref{chemical}) and (\ref{energyc}) to calculate the energy of different temperatures for $\xi=0$ and $\xi=1$ so that one may follow our calculation and method more easily. By reliable interpolation and fitting, we can get two energy functions $f_0(T)$ for $\xi=0$ and $f_1(T)$ for $\xi=1$. For a given $E$, we get numerically $T_0$ and $T_1$ by solving $E=f_0(T)$ and $E=f_1(T)$, respectively. In this case, we can determine the coefficients $a(E)$ and $b(E)$ with Eq. (\ref{RelationXiT}). By solving further the equation $-1=a(E)+b(E)T^2$, we get the corresponding fermion temperature $T(E)$ for this energy. Following this method, we get a series of $E(T)$ for fermions from the information of $f_0(T)$ and $f_1(T)$ without suffering from fermion sign problem. The heat capacity for fermions can be obtained with $C(T)=dE(T)/dT$. In a testing simulation, the heat capacity is more demanding than energy, hence in Fig. \ref{heat}, we give the heat capacity based on our method (red line) and the heat capacity (blue dashed line) directly from Eqs. (\ref{chemical}) and (\ref{energyc}).

\begin{figure}[htbp]
\begin{center}
 \includegraphics[width=0.75\textwidth]{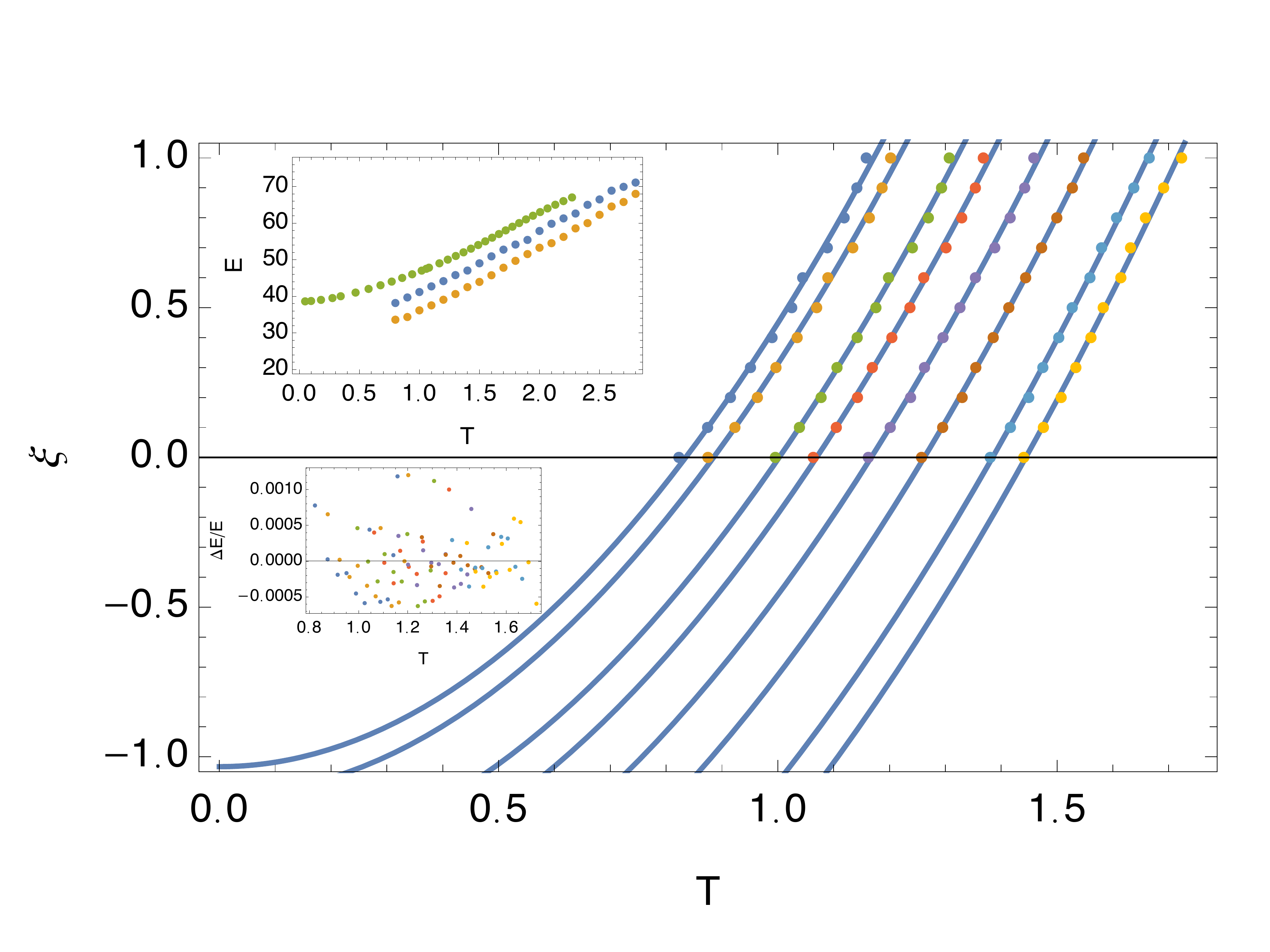} 
\caption{\label{intxi} For 10 interacting particles, the circles from the left to right correspond to energy contours of 38.6, 39.3, 41, 42, 43.5, 45, 47, 48, respectively. The corresponding lines are the fitting with $a+b T^2$. 
The blue circle and yellow circle in the upper inset show the simulated energy for $\xi=0$ and $\xi=1$, while the green circle is the energy of fermions calculated with our method. In the lower inset, we show the relative deviation, compared with the parabolic function. }
\end{center}
\end{figure}

We consider the two-dimensional harmonic trap by including a Coulomb-type interaction:
\begin{equation} 
V_{int}=\sum_{l<j}^N \frac{\lambda}{{|{\textbf r}_l-{\textbf r}_j|}}.
\end{equation}
Here $\lambda$ represents the dimensionless coupling constant of the Coulomb-type interaction. For $10$ particles and $\lambda=0.5$, with the method in Ref. \cite{XiongFSP,XiongPara}, for different temperatures, we obtain energies for $\xi=0$ and $\xi=1$ with PIMD  shown in the upper inset of Fig. \ref{intxi}, which enables us to obtain two functions $f_0(T)$ and $f_1(T)$. From $f_0(T)$, $f_1(T)$ and the relation (\ref{RelationXiT}), we get the energy of the fermions for different temperatures. In Fig. \ref{heat}, the blue line gives the heat capacity of 10 interacting fermions with different temperatures.

In Fig. \ref{intxi}, by calculating more data of the energy for different temperatures and $\xi\geq 0$, we verify again that $\xi_E(T)$  satisfies the simple relation (\ref{RelationXiT}). 
In the lower inset of Fig. \ref{intxi}, we show the relative deviation, compared with the parabolic function. The maximum relative deviation is about $0.1\%$, while the mean relative deviation is smaller than $0.04\%$, which does show good agreement with the parabolic function.
We have also found that the fermion energy ($E=46.83$) for $\beta=1$ does not conflict with the result ($E=49\pm 3$) by Dornheim \cite{Dornheim} and our previous calculation \cite{XiongPara}, while in both Ref. \cite{XiongPara} and Ref.  \cite{Dornheim}, it is difficult to consider the temperature below $\beta=1$ and impossible for $T<<1$. The extrapolation method in Ref. \cite{XiongPara} predicts $E\approx 49.9$, which is larger than the result of this work, because the potential inflection point will make the result of the extrapolation method based on the energy data of $\xi\geq 0$ always larger than the actual energy.

In all our results in this work, the error due to the statistical fluctuations is negligible, hence we do not give the error bar in the present work. Of course, in practical application or precise calculations one may give more accurate calculation and more careful analysis of the statistical fluctuations. It is worth pointing out that the main purpose of the present work is to propose and verify the idea of our method. Hence, we only use moderate $10^7$ MD steps and $P=12/T$ beads with separate Nosé-Hoover thermostat \cite{Nose1,Nose2,Hoover,Martyna,Jang} in our calculation to assure convergence, and to test our idea and also show the efficiency of our method. In practical applications, to satisfy the high precision calculation of some problems, we may consider to increase significantly the MD steps and the number of beads $P$ per particle.

We may improve on this scheme by employing more complicated inference processes, for example by using higher order curves to describe the contours. By satisfying the condition of $\left.\partial \xi(E,T)/\partial T\right|_{T=0}=0$, we may consider the following more general expression:
\begin{equation}
\xi+d(E)\xi^2=a(E)+b(E)T^2+c(E)T^3.
\label{more}
\end{equation}
In this case, to determine the coefficients $a(E),b(E),c(E),d(E)$, we need to calculate $E(\xi\geq 0,T)$ with four different parameters $\xi$. In the appendix, we use this formula to give more accurate calculation of the fermion energy for greater number of particles with and without interparticle interactions. Even for dozens of particles, the constant energy contour remains a concave curve, there is no inflection point, so in principle we can always infer the curve from sufficient data in the $\xi\geq 0$ half plane. However, as the number of particles increases, the contour curve tends to take a sharper turn (larger second derivative), so more data and more parameters are needed to capture this behavior, as shown in the appendix.

It is worth pointing out that, in practice, one can use some reference (experimental data or estimation) to roughly determine the fermionic energy at temperatures of interest. Then we run the simulation for $\xi\geq 0$ on a wide range of temperatures to pin down at which temperature range is the corresponding fermionic energy located. Even if no prior information about fermions is available, we can still get the energies for a range of temperatures by running more simulations on $\xi\geq 0$ half plane.

\section{Conclusions}

As a summary, in this work we considered the three dimensional phase space based on the parametrized path integral formulation \cite{XiongFSP,XiongPara}, and proposed a scheme to infer the fermion energy at all temperatures from the available information in the half space with $\xi\geq0$, via constant-energy contours. 
We successfully applied the present method to study both noninteracting and interacting particles and obtained reasonable heat capacity curve. The scope of application of the present method is well beyond that of traditional method and it is expected this new method can be applied to study some Fermi systems previously intractable by other simulation techniques \cite{nodes,Helium,Militzer,Mak,Blunt,Malone,Schoof1,Schoof2,Schoof3,Yilmaz,PB1,PB2,Joonho,WDM}.

From a rigorous mathematical point of view, we cannot proclaim that fermion sign problem has been solved. However, from a practical point of view, at least for some physical systems, we are able to apply PIMD/PIMC combined with the method here to yield accurate calculation of the energy even when the fermion sign problem is very severe. For the $\xi_E$ orbit, we can first check the noninteracting Fermi system based on analytical formulas. If we can verify that the noninteracting case works, that means our method is applicable for the cases of Fermi system with weakly repulsive/attractive interaction. This implies that this method contains intrinsic value for dilute ultra-cold Fermi atomic gases. In Ref. \cite{XiongCooper}, we obtained the heat capacity peak caused by fermionic pairing for two-component attractive Fermi gases based on our current method.

In this paper, we considered the case of strongly repulsive Coulomb interaction. It is a pity we cannot very rigorously compare our results with other simulation results due to the difficulty of fermion sign problem. In future works, we need to apply our method to realistic physical systems to verify its viability, for example the thermodynamic experiments of many trapped Fermi ions or electrons with repulsive Coulomb interaction. Of course, if the $\xi_E$ orbit when $\xi<0$ exhibits strange behavior in some circumstances, it is possible to give erroneous inferences for the fermionic energies. In that case, we can perhaps try other types of orbits to avoid this problem. Thus we need to be careful when applying this method in practice. The applicability of the method requires dedicate studies in the future, comparison with experiments and other simulations can give us valuable insights.



\begin{acknowledgments}
This work is partly supported by the National Natural Science Foundation of China under grant numbers 11175246, and 11334001. 
\end{acknowledgments}

\textbf{DATA AVAILABILITY}

The data that support the findings of this study are available from the corresponding author upon reasonable request. The code of this study is openly available in GitHub (https://github.com/xiongyunuo/PIMD-Pro-2).

\newpage

\appendix

\textbf{Supplementary Material for "On the thermodynamics of fermions at any temperature based on parametrized partition function"}

\setcounter{equation}{0}
\renewcommand\theequation{A.\arabic{equation}}

\setcounter{section}{0}
\renewcommand\thefigure{A.\arabic{section}}

\setcounter{figure}{0}
\renewcommand\thefigure{A.\arabic{figure}}

\section{The details to obtain relation $\xi_E(T)\approx a(E)+b(E)T^2$ based on general physical consideration}

From the following simple relation proved in the text:
\begin{equation}
\left.\frac{\partial\xi(E,T)}{\partial T}\right|_{T=0}=0,
\label{exact}
\end{equation}
for a given energy, if we expand $\xi_E(T)$ about $T$ with Taylor series, there should be an absence of the linear term. In this case, we have
\begin{equation}
\xi_E(T)=a(E)+b(E)T^2+\sum_{n>2}c_n(E)T^n.
\label{Axirelation}
\end{equation}

At the high temperature limit of $\beta\rightarrow 0$ (or $T\rightarrow \infty$), $E(\beta,\xi)$ should be independent of the quantum statistics parameter $\xi$. In this case, we have
\begin{equation}
E(\beta\rightarrow 0,\xi)=const.
\end{equation}
For small $\beta$, based on the above result and the monotonic behavior of $E(\beta,\xi)$ about $\xi$, for fixed $\beta$, $E(\beta,\xi)$ can be approximated well as a linear function of $\xi$, i.e.,
\begin{equation}
E(\beta,\xi)\approx E(\beta,\xi=0)+d(\beta)\xi.
\end{equation}
For small $\beta$ with linear behavior of $\xi$, $E(T,\xi)$ may be written as
\begin{equation}
E(T,\xi)\approx E(T=0,\xi=0)+(\alpha_1+\alpha_2 T)T+(\gamma_1+\gamma_2 T)\xi.
\end{equation}
At high temperature, it is clear that $|\alpha_2|<<|\alpha_1|$ and $|\gamma_2|<<|\gamma_1|$.
In this case, for a fixed energy $E$, we have
\begin{equation}
\xi_E(T)\approx \frac{\Delta E}{\gamma_1}-\left(\frac{\Delta E\gamma_2}{\gamma_1^2}+\frac{\alpha_1}{\gamma_1}\right)T+\left(\frac{\alpha_1\gamma_2}{\gamma_1^2}-\frac{\alpha_2}{\gamma_1}\right)T^2.
\end{equation}
Here $\Delta E=E-E(T=0,\xi=0)$. This means that at high temperature, the expansion to $T^2$ is a good approximation to $\xi_E(T)$. Usually, the larger the independent variable, the more we need to keep higher-order terms. This means the possibility that the expansion to $T^2$ is a good approximation for the whole temperature region.

The combination of the low temperature and high temperature behavior suggests that $\xi_E(T)$ may be approximated well by the following simple expression verified by the calculation in the text.
\begin{equation}
\xi_E(T)\approx a(E)+b(E)T^2.
\label{ARelationXiT}
\end{equation}

To sum up, the above simple relation originates from two physics: (i) the monotonic behavior of $E(\xi,\beta)$ about $\xi$ and $\beta$; (ii) the exact relation (\ref{exact}).

\section{More general consideration of the orbit $\xi_E(T)$}

\setcounter{equation}{0}
\renewcommand\theequation{B.\arabic{equation}}

\begin{figure}[htbp]
\begin{center}
 \includegraphics[width=0.9\textwidth]{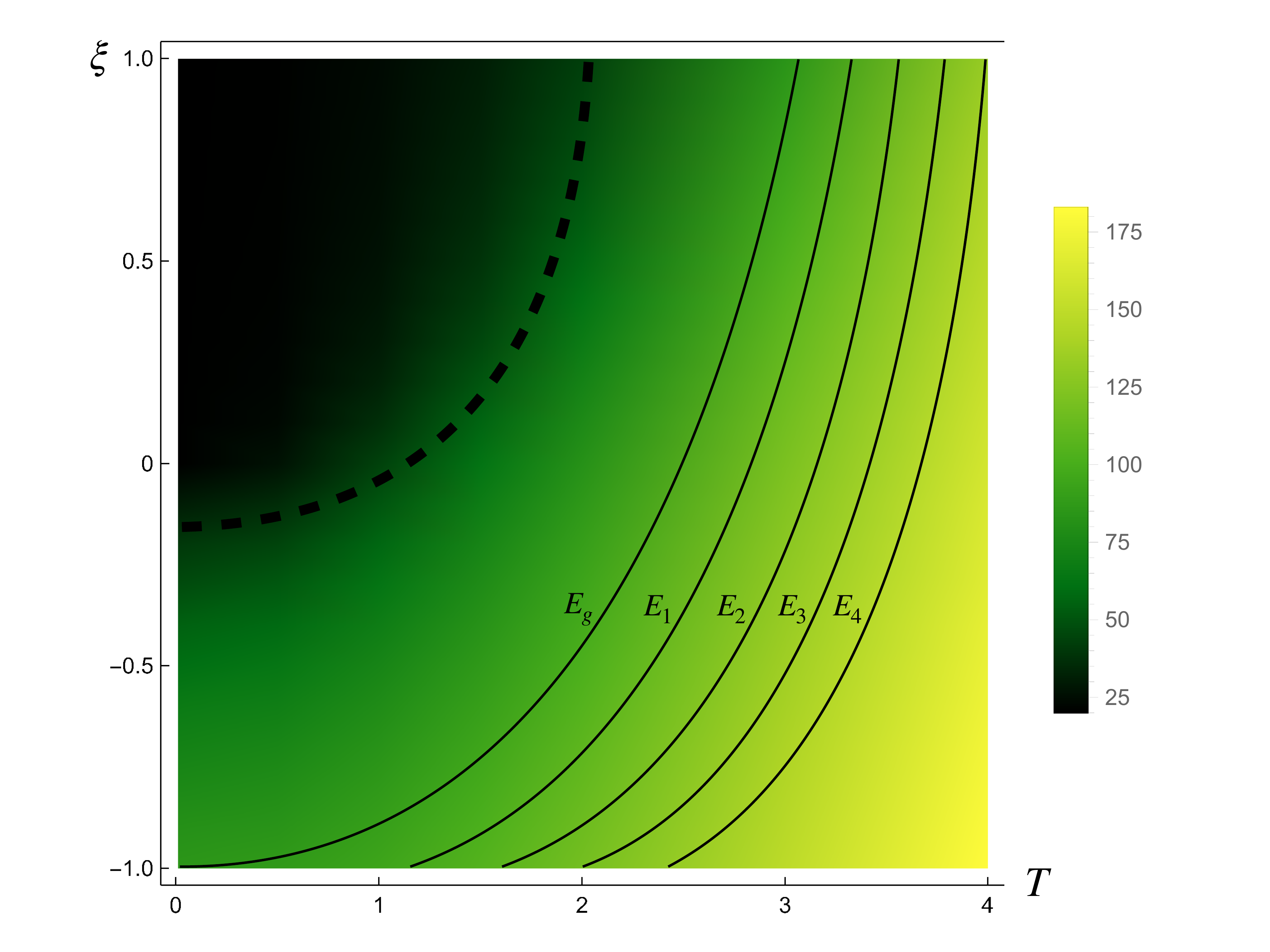} 
\caption{\label{BlackHole} Shown is the complete contour map of 10 noninteracting particles. In the black region on the upper left corner, the energies are almost constant, showing the function $E(\xi,T)$ is non-analytical there for large particle number and this demonstrates why the previous extrapolation scheme fails for this region, in which we attempted to extrapolate along the vertical lines on the contour map. With our new method, however, we bypass this black region and instead do inference along the black curves with constant energy. Those black curves reside in a region where $E(\xi,T)$ has good analytical property so the inference process works for all temperatures.}
\end{center}
\end{figure}

In Fig. \ref{BlackHole}, we illustrate the reason why the present method succeeds to predict the fermion energy from zero temperature to high temperature, while our previous method \cite{XiongFSP,XiongPara} fails for low temperature. In this figure, for 10 noninteracting particles in the two-dimensional harmonic trap, we give the contour map of $E(\xi, T)$. We emphasize that PIMD and PIMC can only give accurate calculation for $\xi\geq 0$. Here we give the full contour map by using the exact expression (\ref{chemical}) and (\ref{energyc}) in the text to show the general structure.

The dashed circle encloses a low-temperature region where $E(\xi,T)$ is almost a constant. For $T=0$ as an example, we will find that $E(\xi\geq 0,T=0)=20$. In this case, it is obvious that we can not predict the fermion energy at zero temperature based on these data along the line of constant temperature. Outside this dark region enclosed by the dashed circle, however, $E(\xi,T)$ has regular change, so that we have the chance to predict the fermion energy from the accurate data of $E(\xi\geq 0,T)$. For each constant $E$, we have a fully defined trajectory $\xi_E(T)$ with the same energy.  Even under this general idea, we still need to have clever method to solve the problem. In hindsight, it is obvious that the orbit $\xi_E(T)$ with constant $E$ gives a good choice to predict the fermion energy. We give a brief reason as follows.

From 
\begin{equation}
E(\xi,T)=constant,
\end{equation}
we see that $\xi$ is a function of $T$ along the constant energy $E$.
In this case, we have
\begin{equation}
\frac{\partial E}{\partial\xi}\frac{d\xi}{dT}+\frac{\partial E}{\partial T}=0.
\end{equation}
This leads to
\begin{equation}
\frac{d\xi}{dT}=-\frac{\partial E/\partial T}{\partial E/\partial \xi}.
\end{equation}
Because $\frac{d\xi}{d T}$ is along the constant energy $E$, we may also write the above equation as
\begin{equation}
\frac{\partial\xi}{\partial T}=-\frac{\partial E/\partial T}{\partial E/\partial \xi}.
\end{equation}
This gives a simple derivation of Eq. (\ref{ppp}) in the text.
From the third law of thermodynamics, we have
\begin{equation}
\left.\frac{\partial E}{\partial T}\right|_{T=0}=0.
\end{equation}
In this case, we have the following exact relation
\begin{equation}
\lim_{T\rightarrow 0}\left.\frac{d\xi}{dT}\right|_{E}=0.
\end{equation}
With the symbol in the text, it is
\begin{equation}
\lim_{T\rightarrow 0}\frac{d\xi_E(T)}{dT}=0.
\end{equation}
It is this exact condition that makes the orbit of constant energy a good choice to predict the fermion energy.

Generally speaking, the orbit satisfying the above condition with constant energy may be written as
\begin{equation}
\xi_E+\sum_{n\geq 2}d_n(E)\xi_E^n=a(E)+b(E)T^2+\sum_{n>2}c_n(E)T^n.
\label{higher}
\end{equation}
All these coefficients are determined by the physics of the system, which is not known in advance. Fortunately, the accurate data of the relation between $\xi\geq 0$ and $T$ for a given constant energy gives us the opportunity to determine accurately these coefficients. The more coefficients we consider, the more accurately we can determine the orbital and thus more accurately obtain the fermion temperature for a given energy. In the main text, we only use two coefficients $a(E)$ and $b(E)$. Now, we consider the following expression which does improve our results, compared with the choice $\xi_E=a(E)+b(E)T^2$ in the text.
\begin{equation}
\xi_E+d(E)\xi_E^2=a(E)+b(E)T^2+c(E)T^3.
\label{four}
\end{equation}
To determine these four coefficients, we need four inputs with $\xi\geq 0$.

In this supplementary material, we use the following program to predict the fermion energy.

(1) For $\xi=0, 0.25, 0.5, 1$, we calculate the energy for different temperatures so that we get four sets of data.

(2) By interpolation and fitting, we get four functions $f_j(T)$ ($j=1,2,3,4$) for different $\xi$.

(3) For a given $E$, by solving $E=f_j(T)$ ($j=1,2,3,4$), we have $\{\xi_j,T_j\}$ ($j=1,2,3,4$).

(4) These solutions $\{\xi_j,T_j\}$ can determine the coefficients $a(E), b(E), c(E), d(E)$ with Eq. (\ref{four}).

(5) By setting $\xi_E=-1$ in Eq. (\ref{four}), we finally get the temperature of the fermions for the given energy.

(6) Repeating the above process, we will get the fermion energy for different temperatures including the zero temperature.

\begin{figure}[htbp]
\begin{center}
 \includegraphics[width=0.9\textwidth]{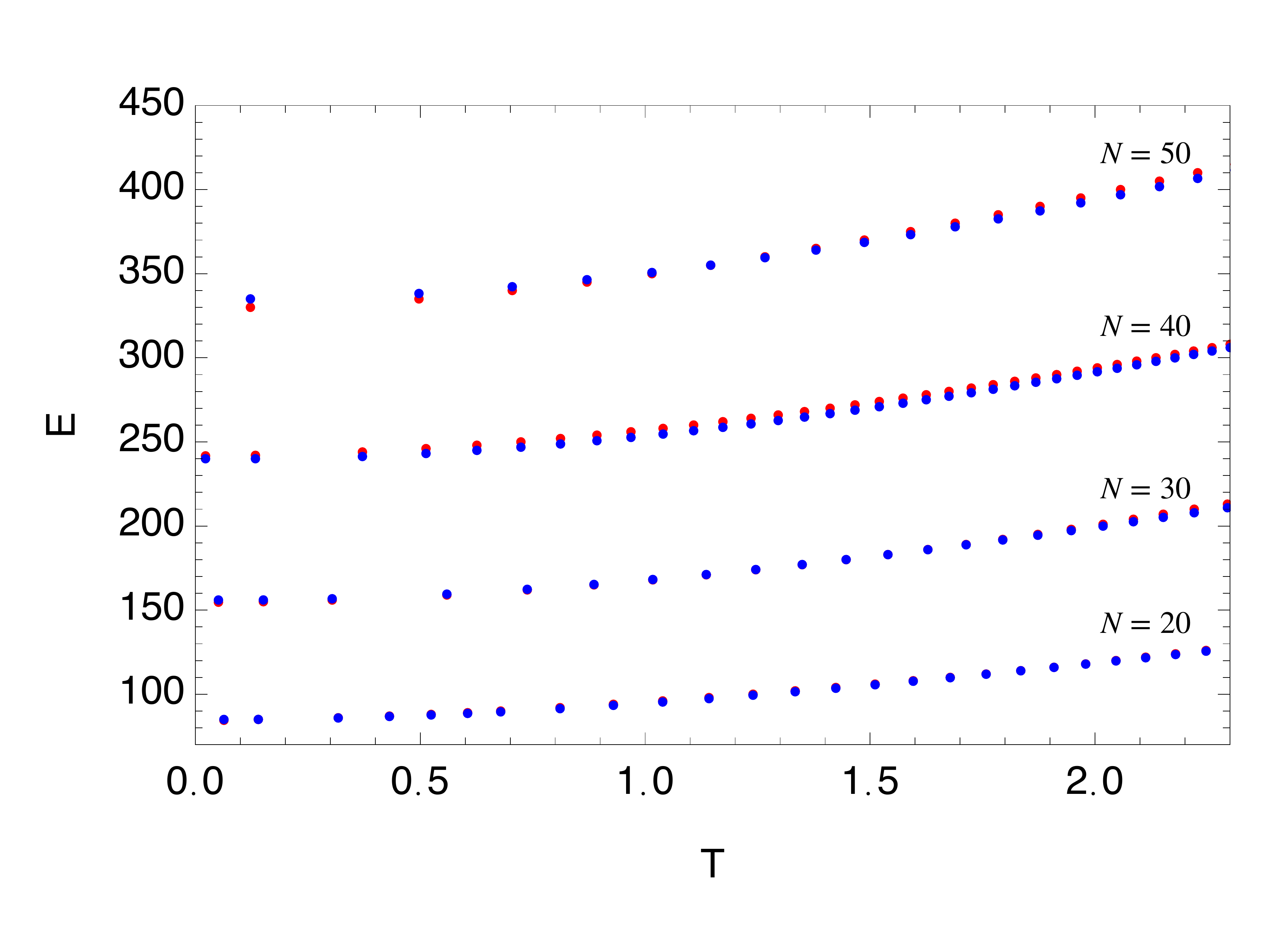} 
\caption{\label{noninteracting}  For particle number $N=20,30,40,50$ without interaction, the red circles show the energy of fermions based on our method, while the blue circles correspond to the result of Eqs. (\ref{chemical}) and (\ref{energyc}) in the text.}
\end{center}
\end{figure}

\begin{figure}[htbp]
\begin{center}
\includegraphics[width=0.9\textwidth]{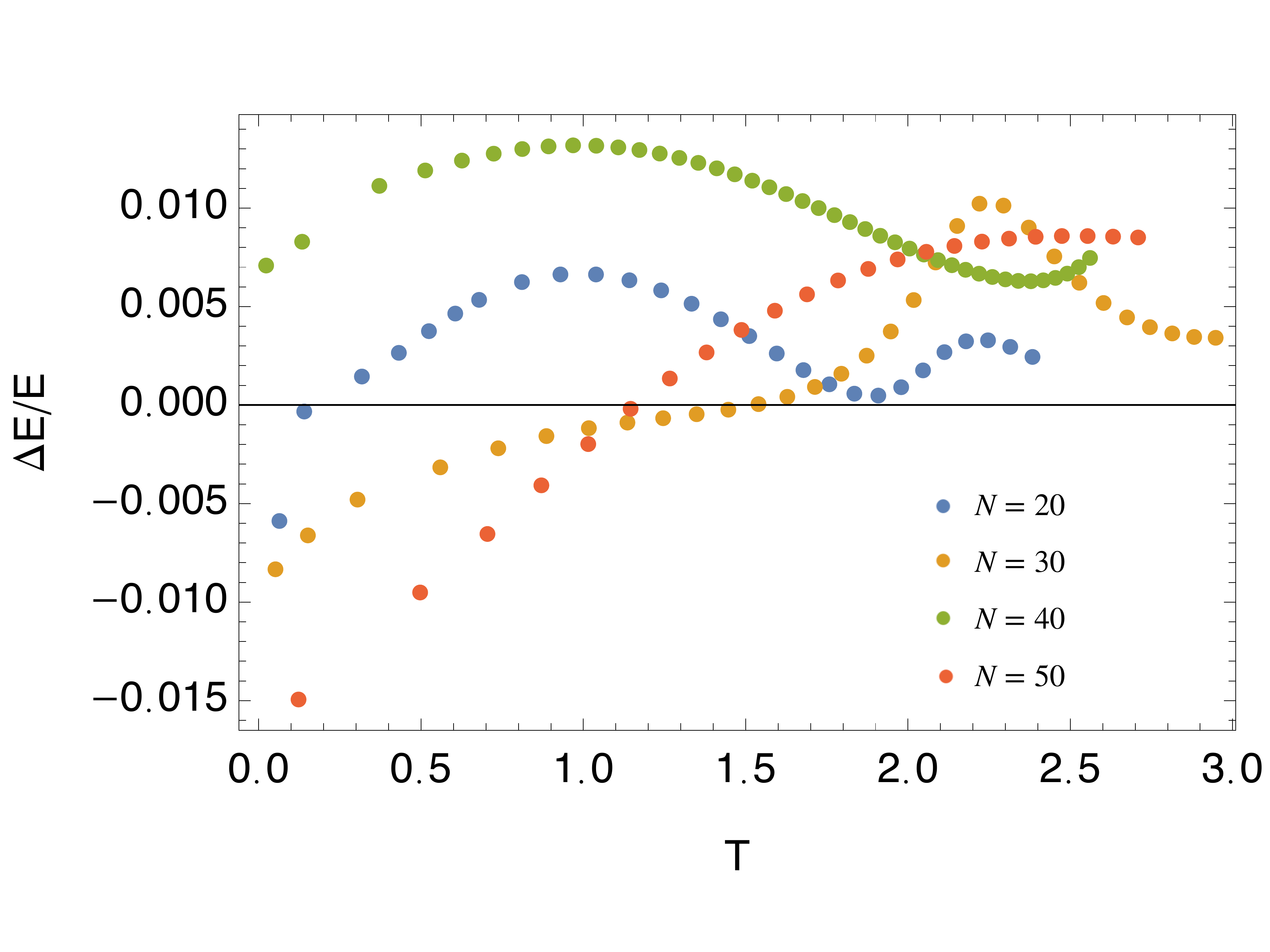} 
\caption{\label{SidealD}  For particle number $N=20,30,40,50$ without interaction, shown are the relative deviations, compared with the analytical result of the energy of fermions.}
\end{center}
\end{figure}

In Fig. \ref{noninteracting}, we give the energy of fermions for different particle number with the above method, and good agreement is found, compared with the results of Eqs. (\ref{chemical}) and (\ref{energyc}). In Fig. \ref{SidealD}, we give the relative deviation $\Delta E/E$, to show clearly the small deviation. Here $\Delta E$ is the difference between our method and the analytical result of Eqs. (\ref{chemical}) and (\ref{energyc}) in the text.

\begin{figure}[htbp]
\begin{center}
 \includegraphics[width=0.9\textwidth]{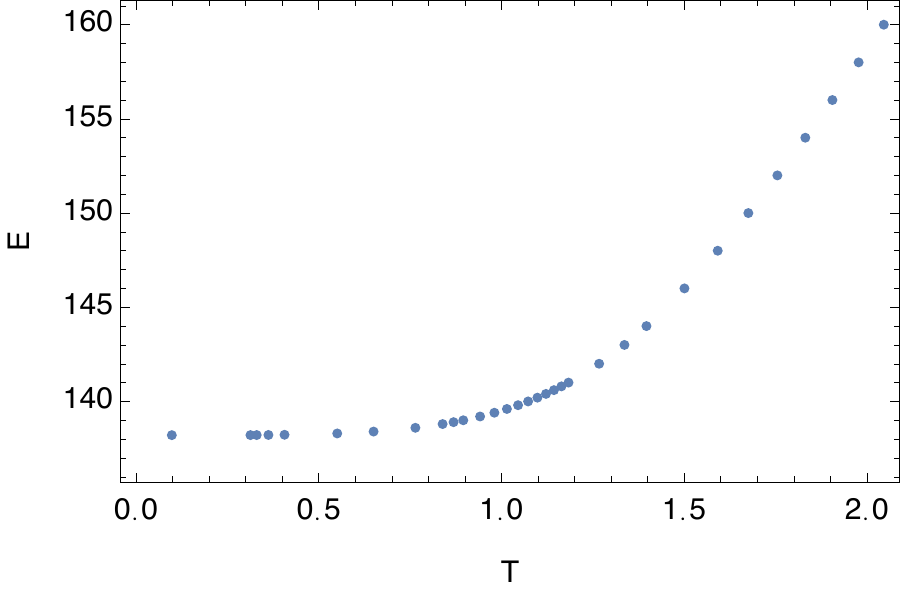} 
\caption{\label{int20}  For particle number $N=20$, shown are the energy of fermions for different temperature with Coulomb interaction of $\lambda=0.5$.}
\end{center}
\end{figure}

In Fig. \ref{int20}, we give the energy of fermions for 20 interacting particles with $\lambda=0.5$.

We emphasize that due to the monotonic behavior of $E(\xi,\beta)$ about $\xi$ and $\beta$ and the exact relation (\ref{exact}), for the examples in this work, Eq. (\ref{ARelationXiT}) is already a good approximation. Hence, higher-order terms in (\ref{higher}) should only give a small correction. This means that the curve $\xi_E(T)$ outside the black region in Fig. \ref{BlackHole} should always show simple behavior, which assures the precision of the thermodynamics of fermions by our method.

\end{document}